\documentclass{PoS}

\usepackage{amsmath}
\usepackage{amssymb}

\newcommand{\bc}{\begin{center}}
\newcommand{\ec}{\end{center}}
\newcommand{\be}{\begin{equation}}
\newcommand{\ee}{\end{equation}}
\newcommand{\bea}{\begin{eqnarray}}
\newcommand{\eea}{\end{eqnarray}}
\newcommand{\bi}{\begin{itemize}}
\newcommand{\ei}{\end{itemize}}
\newcommand{\bt}{\begin{tabular}}
\newcommand{\et}{\end{tabular}}

\def\Nf{N_{\rm f}}

\def\ks{\kappa_\mathrm{sea}}
\def\ksc{\kappa_\mathrm{sea}^c}

\def\mqs{m_{\rm q,sea}}

\def\mps{m_{\rm PS}}
\def\mpss{m_{\rm PS,sea}}
\def\mpsv{m_{\rm PS,val}}
\def\mv{m_{\rm V}}
\def\fps{f_{\rm PS}}

\def\fK{f_{\rm K}}
\def\fv{f_{\rm V}}

\def\fvt{f_{\rm V}^{\perp}}
\def\fvtZ{f_{\rm V}^{\perp(0)}}
\def\fvtO{f_{\rm V}^{\perp(1)}}

\def\fvtRGI{f_{\rm V}^{\perp\,{\rm RGI}}}

\def\fvtMSbar{f_{\rm V}^{\perp\,\overline{MS}}}
\def\ZA{Z_{\rm A}}
\def\cA{c_{\rm A}}
\def\bA{b_{\rm A}}
\def\ZV{Z_{\rm V}}
\def\cV{c_{\rm V}}
\def\bV{b_{\rm V}}
\def\ZT{Z_{\rm T}}
\def\cT{c_{\rm T}}
\def\bT{b_{\rm T}}
\def\fm{\;{\rm fm}}
\def\MeV{\;{\rm MeV}}
\def\GeV{\;{\rm GeV}}

\def\msbar{\overline{\rm MS}}

\def\ks{\kappa_{\rm sea}}
\def\kv{\kappa_{\rm val}}

\PoS{PoS(LAT2005)063}

\title{
\begin{picture}(0,0)(0,0)%
   \put(0,75){\makebox(0,0)[l]{\textnormal{\normalsize DESY 05-182}}}%
   \put(0,60){\makebox(0,0)[l]{\textnormal{\normalsize Edinburgh 2005/11}}}%
   \put(0,45){\makebox(0,0)[l]{\textnormal{\normalsize Liverpool LTH 670}}}%
\end{picture}%
Meson decay constants from $\Nf=2$ clover fermions }

\ShortTitle{Meson decay constants from $\Nf=2$ clover fermions }

\author{
  Meinulf G\"ockeler$^a$,
  Roger Horsley$^b$,
  \speaker{Dirk Pleiter}$^c$,
  Paul~E.L.~Rakow$^d$,
  Gerrit Schierholz$^{c,e}$,
  Wolfram Schroers$^c$,
  Hinnerk St\"uben$^f$ and
  James M.~Zanotti$^c$\\
  \llap{$^a$} Institut f\"ur Theoretische Physik, Universit\"at Regensburg,
              93040 Regensburg, Germany\\
  \llap{$^b$} School of Physics, University of Edinburgh, Edinburgh EH9 3JZ,
              UK\\
  \llap{$^c$} John von Neumann-Institut f\"ur Computing NIC, Deutsches
              Elektronen-Synchrotron DESY, 15739 Zeuthen, Germany\\
  \llap{$^d$} Theoretical Physics Division, Department of Mathematical Sciences,
              University of Liverpool, Liverpool L69 3BX, UK\\
  \llap{$^e$} Deutsches Elektronen-Synchrotron DESY, 22603 Hamburg, Germany\\
  \llap{$^f$} Konrad-Zuse-Zentrum f\"ur Informationstechnik Berlin,
              14195 Berlin, Germany

        E-mail: \email{dirk.pleiter@desy.de}}
\author{QCDSF-UKQCD Collaboration}

\abstract{We present recent results for meson decay constants
calculated on configurations with two flavours of O(a)-improved
Wilson fermions. Non-perturbative renormalisation is applied and
quark mass dependencies as well as finite volume and discretisation
effects are investigated. In this work we also present the first computation
of the coupling of the light vector mesons to the tensor current using
dynamical fermions.
}

\FullConference{XXIIIrd International Symposium on Lattice Field Theory\\
                 25-30 July 2005\\
                 Trinity College, Dublin, Ireland}

\begin{document}

\section{Introduction}

The theoretical description of decay processes which include light
mesons strongly relies on a non-perturbative and model independent
determination of the relevant parameters. The experimentally
most accurately known decay constant is the pseudo-scalar decay constant,
$\fps$, which can be determined from the leptonic decay of charged pions
or kaons. Attempts to compute this quantity from first principles on
the lattice is an important validation of the lattice QCD methodology.
It allows in particular to test control on the required extrapolations
to the infinite volume, continuum and finally chiral limit.
For phenomenology much more interesting is a lattice determination of
the coupling of the light vector mesons to the vector current ($\fv$)
and the tensor current ($\fvt$).
While the constant $\fv$ can in principle be extracted from experimental
studies of, e.g., leptonic decays like $\tau^- \rightarrow V^- \nu_\tau$,
the coupling $\fvt$ can only be determined theoretically.
It has been pointed out by the authors of~\cite{Becirevic:2003pn} that
the form factors describing the semi-leptonic B-decay
$B\rightarrow \rho l \nu$ strongly depend on these couplings. A
precise calculation of these couplings is hence an important step towards
a more precise determination of one of the less well known CKM
parameters, $|V_{\rm ub}|$.

\section{Lattice Details}

The configurations with $\Nf=2$ flavours of non-perturbatively improved
Wilson fermions and Wilson glue which have been generated by the QCDSF
and UKQCD collaborations in recent years now cover a larger range of
quark masses and several lattice spacings. The simulations are
performed for several values of the gauge coupling $\beta$
with the lattice spacing varying between $0.07-0.11 \fm$. For each
lattice spacing configurations are generated for 3-4 sea quark
masses in the range of $590-1170 \MeV$. For additional leverage
on the quark mass dependence we performed simulations at a large number
of valence quark masses which vary between $470$ and $1200 \MeV$.
Also the volumes are varied to explore
finite size effects. Finally, a non-perturbative determination of the
renormalisation constants is performed to further reduce systematic
uncertainties.

Correlation functions for pseudo-scalar and vector mesons are calculated
on configurations taken at a distance of 5-10 trajectories using 8-4
different locations of the fermion source.  We use binning to obtain an
effective distance of 20 trajectories. The size of the bins has little
effect on the error, which indicates auto-correlations are small. To
improve overlap with the ground state we employ Jacobi smearing.

In order to scale our results at different values of $\beta$ and $\ks$
we use the Sommer-scale $r_0$, which can be
determined with good precision on the lattice. When translating this
into more common physical units, we face the problem that the static
quark potential is difficult to determine experimentally. We therefore
choose to use lattice results for $r_0 m_{\rm N}$ (where $m_{\rm N}$ is
the nucleon mass) to obtain $r_0 = 0.467 \fm$.
We typically will use $r_0/a$ at the respective values
of $\beta$ and $\ks$.
Only when stated explicitly we will use $r_0/a$
extrapolated to the chiral
limit~\cite{Gockeler:2005rv}.

To fully eliminate $O(a)$ discretisation errors, it is not sufficient to
improve the action, the operators also need to be improved. To
determine the renormalised pseudo-scalar decay constant we therefore
have to calculate
\be
\fps =
\ZA\;\; (1 + \bA\;a \mqs)\;\;
{\fps^{(0)}} \left(1 + \cA\;\frac{\fps^{(1)}}{\fps^{(0)}}\right),
\ee
where $\ZA$ is the axial vector current renormalisation factor and
$\bA$ a coefficient describing its dependency on the bare sea-quark
mass $a \mqs = \frac{1}{2} (1/\ks - 1/\ksc)$.
$\cA$ is the improvement
coefficient which needs to be chosen such that $O(a)$ discretisation
errors cancel.

\begin{figure}
\bc
\includegraphics[scale=0.5,angle=-90]{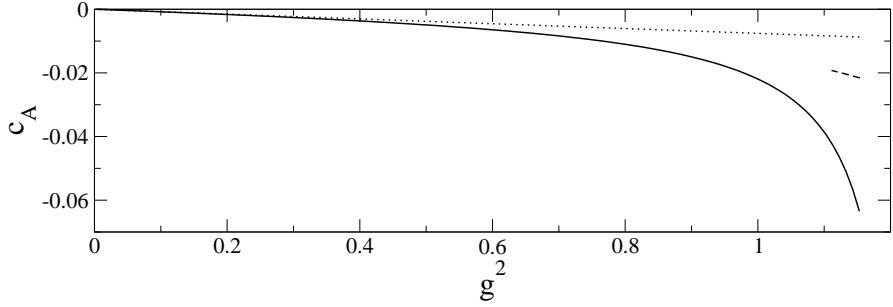}
\vspace*{-5mm}
\ec
\caption{\label{fig:ca}
Comparison of $\cA$ determined from
perturbation theory (dotted line, \cite{sint-weisz}),
boosted perturbation theory (dashed line), and
non-perturbative methods (solid line, \cite{alpha-axial}).}
\end{figure}

The unimproved bare decay constant $\fps^{(0)}$ and the improvement
term are computed from the matrix elements
$\mps \fps^{(0)}   = \langle 0 | A_4 | {\rm PS} \rangle$ and
$\mps a \fps^{(1)} = \langle 0 | a \partial_4 P | {\rm PS} \rangle$.
While the renormalisation constant
$\ZA$ \cite{meinulf-lat05,alpha-axial} (we use~\cite{meinulf-lat05})
and the the critical value of the hopping parameter
$\ksc$ \cite{Gockeler:2004rp}
have been determined non-perturbatively, the coefficient $\bA$ is only known
perturbatively~\cite{sint-weisz}. As the expansion in the bare
gauge coupling $g^2 = 6/\beta$ is known to have bad convergence
properties, we use the boosted coupling $g^{\overline{MS}}(\mu)$ at
the scale $\mu = 1/a$. This coupling is computed from the renormalisation
group equations with
$\mu/\Lambda^{\overline{MS}} = (r_0/a) / (r_0 \Lambda^{\overline{MS}})$,
taking $r_0/a$ in the chiral limit and
$r_0 \Lambda^{\overline{MS}} = 0.617(40)(21)$~\cite{Gockeler:2005rv}.
To assess the differences between perturbation theory, boosted perturbation
theory and non-perturbative methods, we compare the results for $\cA$.
As can been seen from Fig.~\ref{fig:ca} the values from boosted perturbation
theory are much closer to the non-perturbative results. However, there
is still a significant discrepancy between the two results.

Similarly, we define the renormalised and improved vector decay constant
$\fv$:
\be
\frac{1}{\fv} =
\ZV\;\; (1 + \bV\;a \mqs)\;\;
\fv^{(0)} \left(1 + \cV\;\frac{\fv^{(1)}}{\fv^{(0)}}\right),
\ee
where
$e(\lambda)_i\,\mv^2\,\fv^{(0)}
 = \langle 0 | V_i | {\rm V},\lambda \rangle$ and
$e(\lambda)_i\,\mv^2\,a \fv^{(1)}
 = \langle 0 | a \partial_4 T_{i4} | {\rm V},\lambda \rangle$
with $e(\lambda)_i$ being the polarization vector.
The renormalisation constant $\ZV$ and the coefficient $\bV$ have been
computed non-perturbatively~\cite{qcdsf-Z}, but we have to rely on a
perturbative calculation of $\cV$~\cite{sint-weisz}.

Finally, we define the renormalised and improved constant for the coupling
of the vector meson to the tensor current, $\fvt$:
\be
\fvt(\mu) =
\ZT^{\cal S}(\mu)\;\;(1 + \bT\;a \mqs)\;\;
\left(\fvtZ + \cT\;\fvtO\right),
\ee
where
$e(\lambda)_i\,\mv\,\fvtZ
 = \langle 0 | {\rm T}_{\mu\nu} | {\rm V},\lambda \rangle$ and
$e(\lambda)_i\,\mv\,a \fvtO
 = \langle 0 |
    a (\delta_{\mu\rho}\partial_{\rho}V_{\nu} -
       \delta_{\nu\rho}\partial_{\rho}V_{\mu})
    | {\rm V},\lambda \rangle$.
%
For the coefficients $\bT$ and $\cT$ only perturbative results are
available~\cite{sint-weisz}. The QCDSF collaboration has calculated the
renormalisation constant (which depends on both, a scheme $\cal S$ and a scale
$\mu$) non-perturbatively in the $\rm{RI}^\prime$-MOM scheme~\cite{qcdsf-Z}.
We use the 4-loop and 3-loop results for the $\beta$-function and
anomalous dimension to translate our results into the $\msbar$ scheme
at $\mu=2\GeV$~\cite{msbar}.

\begin{figure}[b]
\bc
\includegraphics[scale=0.5,angle=-90]{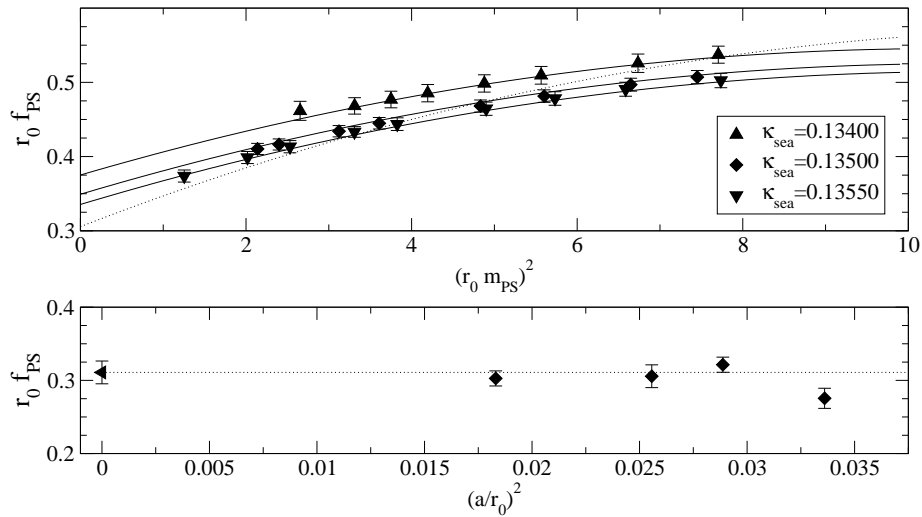}
\vspace*{-5mm}
\ec
\caption{\label{fig:fps}
The upper plot displays the result for $r_0\fps$ at $\beta=5.29$. The lines
show the result of a fit to Eq.~(3.1). Solid lines are
used for the valence quark mass dependency, a dotted for the sea
quark mass dependency.
The lower plot shows $r_0 \fps$ as a function of $(a/r_0)^2$ together
with a fit to a constant. Both $r_0/a$ and $a \fps$ have been extrapolated
to the chiral limit.}
\end{figure}

\section{Results}

\begin{figure}[t]
\bc
\includegraphics[scale=0.5,angle=-90]{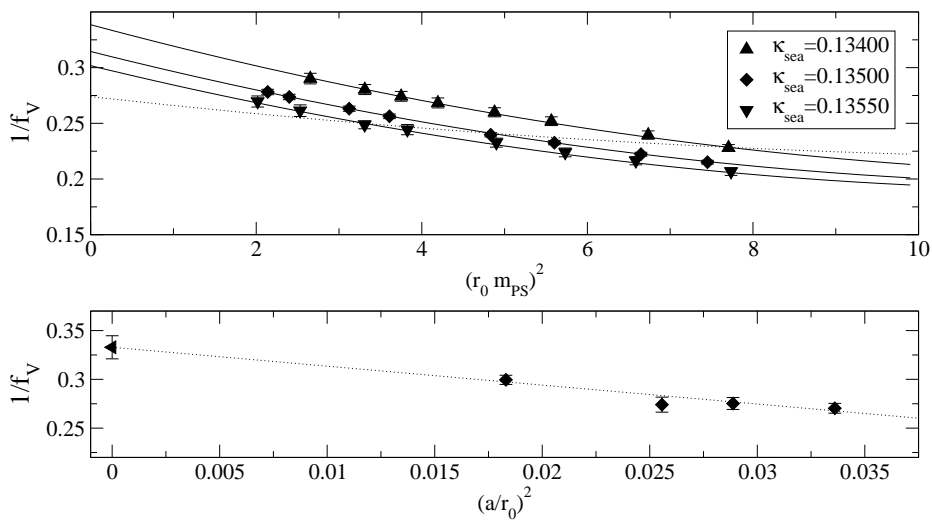}
\vspace*{-5mm}
\ec
\caption{\label{fig:fv}
Similar to Fig.~2, but for $1/\fv$. An ansatz linear in $a^2$
was used for the continuum extrapolation.}
\end{figure}

The results which are obtained for different values of the gauge coupling
$\beta=5.20$, $5.25$, $5.29$ and $5.40$ can be well parametrised by the
following ansatz:
\begin{multline}
r_0 f_x(a,r_0 \mpss, r_0 \mpsv) \quad = \quad 
 r_0 F_x(a)\; \Big[ \;
 1 + \alpha_0\;(r_0\,\mpss)^2 + \alpha_1\;(r_0\,\mpsv)^2 + \\
 \beta_0\;(r_0\,\mpss)^4 + \beta_1\;r_0^4\,\mpss^2\,\mpsv^2 +
  \beta_2\;(r_0\,\mpsv)^4 \;\Big].
\label{eqansatz}
\end{multline}
Here $a \mpss$ denotes the mass of the pseudo-scalar meson for
$\ks = \kv$, while for $a \mpsv$ the valence and sea quark mass might
be different. Given the small number of sea quark masses, we fixed
the value of $\beta_0=0$. In case of $\fvt$ we find the valence
quark mass dependency to be linear in $(a \mpsv)^2$
and therefore we also fix $\beta_2=0$.
The results of our fits to this ansatz are
shown in Fig.~\ref{fig:fps}, \ref{fig:fv} and \ref{fig:fvt}.
As can been seen from these figures, discretisation effects seem to be
small, in particular compared to the statistical errors. For $\fps$ and
$\fvt$ we therefore fit the results in the chiral limit to a constant.
Only in case of $1/\fv$ we find much better agreement when fitting to
a polynomial linear in $(a/r_0)^2$.

For $\beta=5.29$ and $\ks=0.13550, 0.13590$ we determine the decay
constants for three different volumes $L^3\times T=12\times 32$,
$16\times 32$ and $24\times 48$, which corresponds to
$L \approx 1.0, 1.3, 2.0 \fm$.
We find the pseudo-scalar decay constant $\fps$ to be shifted by
$2-6\%$ when increasing the spatial extent from $1.3$ to $2.0 \fm$.
This seems to be larger than one would expect from the results
of chiral perturbation theory~\cite{Colangelo:2005gd}.
No significant finite size effects have been observed for
the vector couplings $\fv$ and $\fvt$.

\begin{figure}[b]
\bc
\includegraphics[scale=0.5,angle=-90]{fvt-b5p29+cont.epsi}
\vspace*{-5mm}
\ec
\caption{\label{fig:fvt}
Similar to Fig.~2, but for $\fvtRGI$.}
\end{figure}

In the chiral limit, our result for the pseudo-scalar decay constant
is $131(7)\MeV$, where only the statistical error has been taken into
account. An extrapolation to the physical pion mass gives a slightly
larger value of $133(7)\MeV$, which still agrees very well with the
experimental result $f_\pi^{\rm exp} = 130.7\MeV$.
Chiral logarithms predicted by chiral perturbation theory, which
have not been taken into account here, would shift this result down.
Extrapolating our results to
the kaon mass gives us an estimate of $\fK \approx 155\MeV$.
This again agrees well with the experimental result
$\fK^{\rm exp} = 159.8\MeV$.
It should be noted however that our estimate for $\fK$ is based
on simulations with degenerate quark masses.
For the rho we find $1/f_\rho = 0.333(12)$. This is slightly larger than
the experimental estimate $1/f_\rho^{\rm exp} \approx 0.27$ obtained
from the branching ratio ${\rm BR}(\tau \rightarrow \rho^-\;\nu_{\tau})$.
Extrapolating our results for the tensor coupling
gives $\fvtRGI = 183(4)\MeV$ or,
after conversion to the $\msbar$-scheme, $\fvtMSbar(2\GeV) = 168(3)\MeV$.
This result is larger compared to previous results from QCD sum
rules~\cite{Ball:1996tb} and from lattice simulations employing the
quenched approximation~\cite{Becirevic:2003pn,fvt-quenched}, which
are in the range of $140-150\MeV$.

\section*{Acknowledgements}

The numerical calculations have been performed on the APEmille at NIC/DESY
(Zeuthen), the Hitachi SR8000 at LRZ (Munich) and on the T3E at EPCC
(Edinburgh)~\cite{ukqcd}. This work is supported in part by the DFG
(Forschergruppe Gitter-Hadronen-Ph\"anomenologie) and by the EU Integrated
Infrastructure Initiative under contract number RII3-CT-2004-506078.
We would like to thank A.~Irving for providing updated results for $r_0/a$
prior to publication.


\begin{thebibliography}{99}

\bibitem{Becirevic:2003pn}
	D.~Becirevic et al.,
	\emph{Coupling of the light vector meson to the vector and to the
	tensor current},
	\emph{JHEP} {\bf 0305} (2003) 007
	[{\tt arXiv:hep-lat/0301020}].

\bibitem{Gockeler:2005rv}
	M.~G\"ockeler et al. [QCDSF Collaboration],
	\emph{A determination of the Lambda parameter from full lattice QCD},
	[{\tt arXiv:hep-ph/0502212}].

\bibitem{meinulf-lat05}
	M.~G\"ockeler et al. [QCDSF Collaboration],
	\emph{The axial charge of the nucleon on the lattice and in chiral
	perturbation theory},
	these proceedings, PoS(LAT2005)349.

\bibitem{alpha-axial}
	M.~Della Morte, R.~Hoffmann and R.~Sommer,
	\emph{Non-perturbative improvement of the axial current for dynamical
	Wilson fermions},
	\emph{JHEP} {\bf 0503} (2005) 029
	[{\tt arXiv:hep-lat/0503003}];
	M.~Della Morte et al. [Alpha Collaboration],
	\emph{Non-perturbative renormalization of the axial current with
	dynamical Wilson fermions},
	\emph{JHEP} {\bf 0507} (2005) 007
	[{\tt arXiv:hep-lat/0505026}].

\bibitem{Gockeler:2004rp}
	M.~G\"ockeler et al. [QCDSF Collaboration],
	\emph{Determination of light and strange quark masses from full
	lattice QCD},
	[{\tt arXiv:hep-ph/0409312}].

\bibitem{sint-weisz}
	S.~Sint and P.~Weisz,
	\emph{Further results on O(a) improved lattice QCD to one-loop order of
	perturbation theory},
	\emph{Nucl.\ Phys.} {\bf B502} (1997) 251
	[{\tt arXiv:hep-lat/9704001}];
	S.~Sint and P.~Weisz,
	\emph{Further one-loop results in O(a) improved lattice QCD},
	\emph{Nucl.\ Phys.\ Proc.\ Suppl.} {\bf 63} (1998) 856
	[{\tt arXiv:hep-lat/9709096}].

\bibitem{qcdsf-Z}
	T.~Bakeyev et al. [QCDSF-UKQCD Collaboration],
	\emph{Non-perturbative renormalisation and improvement of the local
	vector current for quenched and unquenched Wilson fermions},
	\emph{Phys.\ Lett.} {\bf B 580} (2004) 197
	[{\tt arXiv:hep-lat/0305014}];
	M.~G\"ockeler et al. [QCDSF Collaboration], in preparation.

\bibitem{msbar}
	J.~A.~M.~Vermaseren, S.~A.~Larin and T.~van Ritbergen,
	\emph{The 4-loop quark mass anomalous dimension and the invariant
	quark  mass},
	\emph{Phys.\ Lett.} {\bf B 405} (1997) 327;
	J.~A.~Gracey,
	\emph{Three loop $\overline{MS}$ tensor current anomalous dimension
	in QCD},
	\emph{Phys.\ Lett.} {\bf B 488} (2000) 175.

\bibitem{Colangelo:2005gd}
	G.~Colangelo, S.~D\"urr and C.~Haefeli,
	\emph{Finite volume effects for meson masses and decay constants},
	\emph{Nucl.\ Phys.} {\bf B721} (2005) 136
	[{\tt arXiv:hep-lat/0503014}].

\bibitem{Ball:1996tb}
	P.~Ball and V.~M.~Braun,
	\emph{The $\rho$ Meson Light-Cone Distribution Amplitudes of Leading
	Twist Revisited},
	\emph{Phys.\ Rev.} {\bf D54} (1996) 2182
	[{\tt arXiv:hep-ph/9602323}].

\bibitem{fvt-quenched}
	S.~Capitani et al. [QCDSF Collaboration],
	\emph{Towards a lattice calculation of $\Delta(q)$ and $\delta(q)$},
	\emph{Nucl.\ Phys.\ Proc.\ Suppl.} {\bf 79} (1999) 548
	[{\tt arXiv:hep-ph/9905573}];
	G\"ockeler et al. [QCDSF Collaboration], in preparation;
	V.~M.~Braun et al.,
	\emph{A lattice calculation of vector meson couplings to the vector and
	tensor currents using chirally improved fermions},
	\emph{Phys.~Rev.} {\bf D68} (2003) 054501
	[{\tt arXiv:hep-lat/0306006}].

\bibitem{ukqcd}
	C.R.~Allton et al. [UKQCD Collaboration],
	\emph{Effects of non-perturbatively improved dynamical fermions
	in QCD at fixed lattice spacing},
	\emph{Phys.~Rev.} {\bf D65} (2002) 054502
	[{\tt hep-lat/0107021}].

\end{thebibliography}
\end{document}